# Line-Torus Intersection for Ray Tracing: Alternative Formulations

VACLAV SKALA
Department of Computer Science and Engineering
University of West Bohemia
Univerzitni 8, CZ 30614 Plzen
CZECH REPUBLIC
http://www.VaclavSkala.eu

*Abstract:* - Intersection algorithms are very important in computation of geometrical problems. Algorithms for a line intersection with linear or quadratic surfaces are quite efficient. However, algorithms for a line intersection with other surfaces are more complex and time consuming. In this case the object is usually closed into a simple bounding volume to speed up the cases when the given line cannot intersect the given object.

In this paper new formulations of the line-torus intersection problem are given and new specification of the bounding volume for a torus is given as well. The presented approach is based on an idea of a line intersection with an envelope of rotating sphere that forms a torus. Due to this approach new bounding volume can be formulated which is more effective as it enables to detect cases when the line passes the "hole" of a torus, too.

## 1 Introduction

Intersection algorithms play a significant role in all geometric problems and CAD/CAM systems. Intersection algorithms are well documented for linear cases, e.g. line-plane or line-triangle etc., and also for some specific non-linear surfaces like line-sphere intersection etc. However, there are other objects like bicubic parametric patches, torus etc. In this case computation of intersection points is more complex and usually complex formula or iterative formula are to be used.

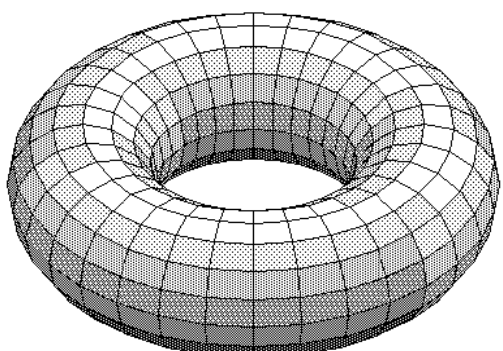

Figure 1: Torus
(Courtesy of Wikipedia)

Intersection of a line and closed surface can be considered as generalized well known clipping problem. Intersection of a line or ray with a surface is the key problem solved in all ray-tracing techniques. Due to the computational complexity a bounding volumes are used to detect cases when a line cannot intersect the given object.

In this paper we present torus-line intersection problem [1] [2], which leads to a quartic equation [3] in principle, and show other possible formulations of the line-torus intersection problem which offer quite different representations of the problem. These reformulations lead to a formulation of a new problem – generalized line clipping by an envelope (convex or non-convex) of parametric closed surfaces.

## 2 Torus Line Intersection

Torus-line intersection is actually a solution of a line in E$^3$ usually given in a parametric form as

$$\boldsymbol{x}(t) = \boldsymbol{x}_A + \boldsymbol{s}\, t$$
$$\boldsymbol{s} = [s_x, s_y, s_z]^T \qquad (1)$$

and a torus, which is generally a surface of the 4$^{th}$ order and can be given as :

$$(x^2 + y^2 + z^2 + R^2 - r^2)^2 = 4R^2(x^2 + y^2) \qquad (2)$$

An alternative formulation

$$\left(R - \sqrt{x^2 + z^2}\right)^2 + y^2 = r^2 \qquad (3)$$

Note that the $z$ axis is the rotational axis. The torus equation can be reformulated as

$$(\boldsymbol{x}^T \boldsymbol{x} + \xi)^2 = 4R \boldsymbol{x}^T \boldsymbol{M} \boldsymbol{x}$$
$$\boldsymbol{x} = [x, y, z]^T \qquad (4)$$

where

$$\xi = R^2 - r^2 \qquad \boldsymbol{M} = \begin{bmatrix} 1 & 0 & 0 \\ 0 & 1 & 0 \\ 0 & 0 & 0 \end{bmatrix} \qquad (5)$$

As there will be some geometric transformations used latter on we can also scale the given torus and a line so that $R = 1$, i.e. the torus is "normalized".

Now the intersection of a line and the torus is given as a solution of equations:
$$\boldsymbol{x}(t) = \boldsymbol{x}_A + \boldsymbol{s}\, t \tag{6}$$
and
$$(\boldsymbol{x}^T\boldsymbol{x} + \xi)^2 = 4\boldsymbol{x}^T\boldsymbol{M}\boldsymbol{x} \tag{7}$$

Substituting Eq.5 to Eq.6 we get

$$[(\boldsymbol{s}t + \boldsymbol{x}_A)^T(\boldsymbol{s}t + \boldsymbol{x}_A) + \xi]^2 = 4(\boldsymbol{s}t + \boldsymbol{x}_A)^T\boldsymbol{M}(\boldsymbol{s}t + \boldsymbol{x}_A) \tag{8}$$

and finally we get
$$[\boldsymbol{s}^T\boldsymbol{s}\, t^2 + 2\boldsymbol{s}^T\boldsymbol{x}_A\, t + \boldsymbol{x}_A^T\boldsymbol{x}_A + \xi]^2$$
$$-4[\boldsymbol{s}^T\boldsymbol{M}\boldsymbol{s}\, t^2 + 2\boldsymbol{s}^T\boldsymbol{M}\boldsymbol{x}_A\, t \tag{9}$$
$$+ \boldsymbol{x}_A^T\boldsymbol{M}\boldsymbol{x}_A] = 0$$

This equations is quite complex, but by detailed evaluation we get a quartic equation
$$at^4 + bt^3 + ct^2 + dt + e = 0 \tag{10}$$
where:
$$\alpha = \boldsymbol{s}^T\boldsymbol{s} \qquad \beta = \boldsymbol{s}^T\boldsymbol{x}_A$$
$$\gamma = \boldsymbol{x}_A^T\boldsymbol{x}_A \qquad \delta = (\gamma + R^2 - r^2)$$
$$\sigma = (\gamma - R^2 - r^2)$$

$$a = \alpha^2 \qquad b = 4\alpha\beta$$
$$c = 2\alpha(\gamma + R^2 - r^2)$$
$$\quad - 4R^2(s_x^2 + s_y^2) + 4\beta^2 \tag{11}$$
$$d = 8R^2 z_A s_z + 4\beta\sigma$$
$$e = \alpha^2 + (R^2 - r^2)^2$$
$$+ 2\begin{bmatrix} s_x^2 s_y^2 + s_z^2(R^2 - r^2) \\ +(s_x^2 + s_y^2)(s_z^2 - R^2 - r^2) \end{bmatrix}$$

It can be seen that the computation can be simplified for the case, when $\boldsymbol{s}^T\boldsymbol{s} = 1$, i.e. the directional vector of the line is normalized or the equation is divided by $a$.

It means that we are getting a quartic equation in the from [4]
$$t^4 + bt^3 + ct^2 + dt + e = 0 \tag{12}$$
which can be simplified by substitution
$$t = x - \frac{b}{4} \tag{13}$$
to
$$x^3 + px^2 + qx + r = 0 \tag{14}$$
where
$$p = \frac{3}{8}b^2 + c$$
$$q = \frac{1}{8}b^3 - \frac{1}{2}bc + d \tag{15}$$
$$r = -\frac{3}{256}b^4 + \frac{1}{16}b^2 c - \frac{1}{14}bd + e$$

If solution for $x$ is found, then the solution of the original equation is given by Eq.12. To get a solution for $x$ the following a qubic equation has to be solved
$$\xi^3 - \frac{p}{2}\xi^2 - r\xi + \frac{4rp - q^2}{8} = 0 \tag{16}$$
Then the $x$ values can be computed from real solution of the equation above as two quadratic equations as follows:
If $q \geq 0$ then
$$x^2 + x\sqrt{2\xi - p} + \xi - \sqrt{\xi^2 - r} = 0$$
$$x^2 - x\sqrt{2\xi - p} + \xi + \sqrt{\xi^2 - r} = 0 \tag{17}$$
If $q < 0$ then
$$x^2 + x\sqrt{2\xi - p} + \xi + \sqrt{\xi^2 - r} = 0$$
$$x^2 - x\sqrt{2\xi - p} + \xi - \sqrt{\xi^2 - r} = 0 \tag{18}$$
It can be seen that the solution itself is not simple, but the formula is closed.

On the opposite, an iterative method like Bisection or Newton method can be used. However there are up to 4 intersections of the line and the torus, so it is necessary to find relevant intervals for $t$, with one intersection only.

### 2.1 Alternative Torus Representation
There are other formulations of a torus as follows, but they are not convenient for our purposes.
$$\left(R - \sqrt{x^2 + y^2}\right)^2 + z^2 = r^2 \tag{19}$$
or a parametric form as
$$x(\varphi, \vartheta) = (R + r\cos\vartheta)\cos\varphi$$
$$y(\varphi, \vartheta) = (R + r\cos\vartheta)\sin\varphi \tag{20}$$
$$z(\varphi, \vartheta) = r\sin\vartheta$$
It can be seen that a solution of a line-torus intersection is not a simple task and it leads to a non-trivial computational problem.

However, there are some other geometrically equivalent formulations which could be used for finding a solution. In the following we will consider only circular torus.

### 2.2 Geometric Transformations
Geometric transformations with points are defined in the projective space using homogeneous coordinates, i.e. in the projective extension of the Euclidean space. A point $\boldsymbol{X} = (X, Y)$ in the Euclidean coordinates has homogeneous coordinates $\boldsymbol{x} = [x, y: w]^T$; $w$ is the homogeneous coordinate. The conversion between the projective space and the Euclidean space is defined as
$$X = \frac{x}{w} \qquad y = \frac{x}{w} \qquad w \neq 0 \tag{21}$$
It means that the projective representation is actually a one parametric set. A point in the Euclidean space $E^2$ is represented as a line with the origin of the coordinate system excluded in the projective space. Geometric transformations with points like rotation, translation, mirroring etc. can be than described by the $Q$ matrix as
$$x' = Qx \quad (22)$$
Note that $x, y$ might have some physical meaning and units, e.g. [m], while $w$ has no unit, it is just a "scaling factor". That's why we used ":" to separate the values in the vector notation.

A line in $E^2$ determined by two points given in the homogeneous coordinates can be computed using the cross product as [5], [6].
$$p = x_1 \times x_2 = \begin{bmatrix} i & j & k \\ x_1 & y_1 & w_1 \\ x_2 & y_2 & w_2 \end{bmatrix} \quad (23)$$
$$p = [a, b: c]^T$$
$$p: ax + by + cw = 0$$
Intersection of two lines $p_1$ and $p_2$ in $E^2$ can be computes as
$$x = p_1 \times p_2 = \begin{bmatrix} i & j & k \\ a_1 & b_1 & c_1 \\ a_2 & b_2 & c_2 \end{bmatrix} \quad (24)$$
$$x = [x, y: w]^T$$
We can see that both computations are in the $E^2$ case "dual", i.e. line and points are dual [7]. In the $E^3$ case a point is dual to a plane and vice versa. It can be shown that a plane given by three points can be determined by the extended cross product as
$$\rho = x_1 \times x_2 \times x_3$$
$$= \begin{bmatrix} i & j & k & l \\ x_1 & y_1 & z_1 & w_1 \\ x_2 & y_2 & z_2 & w_2 \\ x_3 & y_3 & z_3 & w_3 \end{bmatrix} \quad (25)$$
$$\rho = [a, b, c: d]^T$$
$$\rho: ax + by + cz + dw = 0$$
Again, an intersection of three planes can be computed as, see [7], [8], [9] for details
$$x = \rho_1 \times \rho_2 \times \rho_3$$
$$= \begin{bmatrix} i & j & k & l \\ a_1 & b_1 & c_1 & d_1 \\ a_2 & b_2 & c_2 & d_2 \\ a_3 & b_3 & c_3 & d_3 \end{bmatrix} \quad (26)$$
$$x = [x, y, z: w]^T$$
This approach is simple, easy to implement and convenient for GPU implementation as well.

However, matrix transformations for points cannot be used for geometric transformations with lines in the $E^2$ case nor with planes in the $E^3$ case. It can be shown [6] that if a line $p$ is given by two points $x_1$ and $x_2$ and those points are geometrically transformed using the $T$ matrix, i.e.
$$p = x_1 \times x_2 \quad (27)$$
and
$$p' = x'_1 \times x'_2 = Tx_1 \times Tx_2 \quad (28)$$

then
$$p' = Q(x_1 \times x_2) \quad (29)$$
It can be shown that the matrix $Q$ is defined as
$$Q = det(T)(T^{-1})^T \quad (30)$$
Because $p'$ are coefficients of an implicit equation we can simply write
$$p' = det(T)(T^{-1})^T(x_1 \times x_2) \quad (31)$$
As the implicit form is used, coefficients of a line can be multiplied by any non-zero constant and the line will be same. Therefore
$$p' \triangleq (T^{-1})^T(x_1 \times x_2) \quad (32)$$
where $\triangleq$ means protectively equal. Similarly for a plane $\rho$
$$\rho' \triangleq (T^{-1})^T(x_1 \times x_2 \times x_3) \quad (33)$$
It means that we can correctly manipulate with lines and planes, now.

### 2.3 Bounding Volume
Let us assume that the torus lies in the $x - z$ plane, i.e. the $y$-axis is its rotational axis. Bounding volume, defined in [1], is based on an idea that torus is bounded by an intersection of a sphere and two half-spaces, Fig.2.

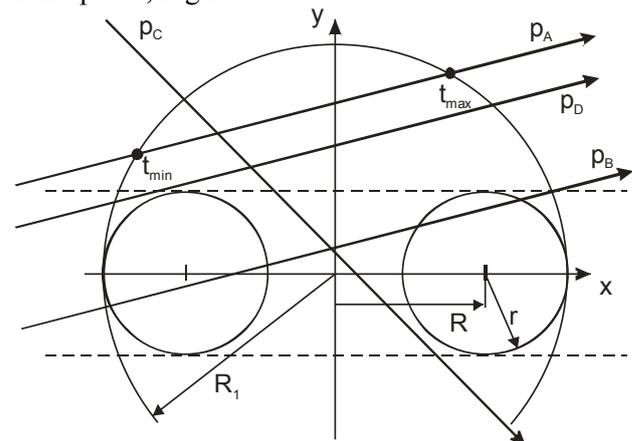

Figure 2: Bounding volume

The radius of the enclosing sphere is given as
$$R_1 = R + r \quad (34)$$
The bounding test computes intersection of a line with a sphere. If such intersections $t_{min}$ and $t_{min}$ exist then the line does not intersect the torus if the following condition is valid [1]
$$\begin{aligned} y_{min} &= y_A + s_y t_{min} \\ y_{max} &= y_A + s_y t_{max} \\ (y_{min} &> r \text{ } \boldsymbol{and} \text{ } y_{max} > r) \text{ } \boldsymbol{or} \\ (y_{min} &< r \text{ } \boldsymbol{and} \text{ } y_{max} < r) \end{aligned} \quad (35)$$
It can be seen that the test does not eliminate cases when a line:
- is passing the "hole inside of the torus" without touching or intersecting the torus – line $p_C$
- nearly touches the torus – line $p_D$ – but there is a small probability

It should be noted that the Fig.2 presents general situation in the $E^3$ case.

### 2.4 Torus Transformation
So far we have dealt with a general situation expecting that the torus is in its basic position, i.e. it lies in the $x - z$ plane and the $y$ axis is the rotational axis. In the case of torus general position the following transformations can be used:

$$Q = \begin{bmatrix} 0 & u & 0 & 0 \\ 0 & n & 0 & 0 \\ 0 & u \times n & 0 & 0 \\ -c_x & -c_y & -c_z & 1 \end{bmatrix} \quad (36)$$

where: $u$ defines $x$ axis of the torus, $n$ defines $y$ axis of the torus, $u \times n$ is used to get an orthonormal basis, and $c$ is the torus centre.

It can be seen that there are some interesting properties of the line-torus intersection problem, like
- torus rotational symmetry,
- if mirroring operation is used only one quadrant can be considered to solve the intersection problem.

We will explore if those properties can contribute to simplification of computation in the following part.

### 2.5 Intersections Classification
As a torus is rotationally invariant we can rotate the given line about $y$ axis so that it lies in a plane $z = z_c$, i.e. in a plane parallel to the $x - y$ plane. There is no significant computational expense as the transformation matrix is accumulated with the $Q$ matrix. Now we can distinguish three fundamentally different cases according to the $z_c$ value:
a) $z_c < R - r$: generally intersection with two independent parts have to be considered, i.e. for $x > 0$ and $x < 0$ and due to convexity each part could have up to 2 intersections only (2 convex envelopes are generated),
b) $R - r \leq z_c < R$: this case is more complex as the envelope has only one part, but it is not convex as it can have an inflexion point and 3 intersection points can be generated,
c) $R \leq z_c < R + r$: the simplest case as only one convex envelope is generated.

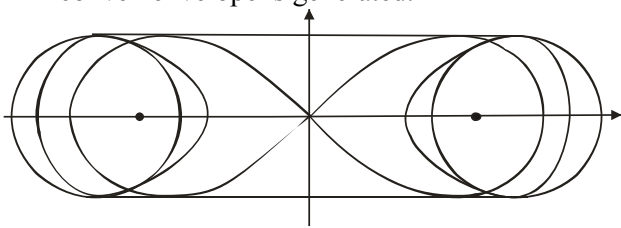

Figure 3: Torus plane intersection for $z_c \leq R - r$

The above mentioned three cases differ significantly. Unfortunately the envelope is not convex in all the cases.

### 2.6 Vieta's Formula
Let us assume that $P(x)$ is a polynomial of degree $n$
$$P(x) = a_n x^n + a_{n-1} x^{n-1} + \cdots \\ + a_1 x + a_0 \quad (37)$$
Then according to the Vieta's formula the roots satisfy equations
$$x_1 + x_1 + \cdots + x_1 = -\frac{a_{n-1}}{a_n}$$
$$(x_1 x_2 + x_1 x_3 + \cdots + x_1 x_n) + \\ (x_2 x_3 + \cdots + x_2 x_n) + \cdots + \\ x_{n-1} x_n = \frac{a_{n-1}}{a_n} \quad (38)$$
$$\vdots$$
$$x_1 x_2 x_3 \ldots x_n = (-1)^n \frac{a_0}{a_n}$$

In the quadratic equation case
$$at^2 + bt + c = 0 \quad (39)$$
we obtain
$$t_1 + t_2 = -\frac{b}{a} \qquad t_1 t_2 = \frac{c}{a} \quad (40)$$
These formulas are not well known and will be used latter on. In the following we will show different approaches to the line – torus intersection problem.

## 3 New Intersection Formulations
In the previous part we have presented the "traditional" approach to the line–torus intersection detection and computation. Now, different equivalent formulations, which could lead to simpler and faster solutions, will be formulated in the following part. They can be briefly classified as follows:
- a sphere is rotating about $y$ axis (the envelope forms a torus) and intersection with the line in $E^3$ is computed directly,
- a sphere is fixed on the $x$ axis and intersection with the line rotating about $y$ axis (i.e. it is actually an intersection of a sphere and double cone) in $E^3$ is computed directly,
- a sphere is rotating about $y$ axis and intersection with the plane $z = z_c$ in $E^3$, on which the given line lies, results into circles in this plane, i.e. circles in $E^2$, forming an envelope, i.e. a curve is given as an intersection of a torus with a plane. An intersection of the envelope of all circles and the line is computed in $E^2$.

  This is actually a generalized line-clipping problem.

Let us explore the first possible formulation more in detail, now.

### 3.1 Sphere Rotation - Intersection in E³

Let us consider a situation in which a torus and line are in the same relative position, but using the above mentioned geometric transformation, the torus is in its basic position, i.e. in the $x - z$ plane.

A torus can be represented as a union all spheres with a radius $r$ rotating about $y$ axis in the $x - z$ plane with a radius $R$. It means that the torus can be defined as a union, i.e. an envelope, of all rotating spheres about $y$ axis as

$$\bigcup (x - x_s(\varphi))^T (x - x_s(\varphi)) - r^2 = 0 \quad (41)$$

$$x(t) = x_A + s\, t$$

where: $\varphi \in\, <0, 2\pi)$, $x_s(\varphi) = R[cos\varphi, 0, sin\varphi]^T$

Now the problem line-torus intersection is transformed to a generalized line clipping problem, when a line is clipped by an envelope of all rotating spheres K which forms the T torus, i.e.

$$T = \bigcup_{\varphi \in <0,2\pi)} K(\varphi, R, r) \quad (42)$$

where $R$ and $r$ are given constants of the torus.

Due to the rotational symmetry about the $y$ axis, the torus and the line can be rotated about $y$ axis so that the line will lie in a plane parallel to the $x - y$ plane.

Now, the given line is defined as

$$x(t) = x_A + s\, t \quad (43)$$

A point $x_A$ and a directional vector $s$ of the line are defined as

$$x_A = [x_A, y_A, z_C]^T \quad s = [s_x, s_y, 0]^T \quad (44)$$

where: $z_C = z_A$ as the line lies in a plane parallel to the $x - y$ plane, i.e. $\rho$: $z = z_C$.

The problem of a line-torus intersection problem is transformed to generalized line clipping problem in E² actually, when a line is clipped by a parametric envelope.

A line is given in the case of E³ as

$$x(t) = x_A + s\, t \quad (45)$$

and a sphere

$$(x - x_s(\varphi))^T (x - x_s(\varphi)) - r^2 = 0 \quad (46)$$

substituting we get

$$[st + \xi(\varphi)]^T [st + \xi(\varphi)] - r^2 = 0 \quad (47)$$
$$\xi(\varphi) = x_A - x_s(\varphi)$$

i.e.

$$s^T s\, t^2 + 2 s^T \xi(\varphi) t + \xi^T(\varphi)\xi(\varphi) - r^2 = 0 \quad (48)$$

where

$$s^T \xi(\varphi) = s^T x_A - s^T x_s(\varphi)$$
$$= [s_x, s_y, 0]\{[x_A, y_A, z_C]^T \quad (49)$$
$$- R[cos\varphi, 0, sin\varphi]^T\}$$

$$= s_x x_A + s_y y_A - R s_x cos\varphi$$

$$\xi^T(\varphi)\xi(\varphi) = x_A^T x_A - 2 x_A^T x_s(\varphi) \quad (50)$$
$$+ x_s^T(\varphi) x_s(\varphi)$$

where:

$$x_s^T(\varphi) x_s(\varphi) \quad (51)$$
$$= R^2 [cos\varphi, 0, sin\varphi][cos\varphi, 0, sin\varphi]^T$$
$$= R^2 (cos^2\varphi + sin^2\varphi) = R^2$$

and

$$x_A^T x_A = x_A^2 + y_A^2 + z_C^2 \quad (52)$$

and

$$x_A^T x_s(\varphi)$$
$$= R[x_A, y_A, z_C][cos\varphi, 0, sin\varphi]^T \quad (53)$$
$$= R(x_A cos\varphi + z_C sin\varphi)$$

Therefore

$$\xi^T(\varphi)\xi(\varphi) = \quad (54)$$
$$x_A^2 + y_A^2 + z_C^2 - 2R(x_A cos\varphi + z_C sin\varphi) + R^2$$

The quadratic equation is now

$$s^T s\, t^2 + 2(s_x x_A + s_y y_A$$
$$- s_x R cos\varphi)t$$
$$+ x_A^2 + y_A^2 + z_C^2 \quad (55)$$
$$- 2R(x_A cos\varphi + z_C sin\varphi) + R^2$$
$$- r^2 = 0$$

In the case of the normalized directional vector $s$, i.e. $\|s\| = 1$, resp. $s^T s = 1$, we get a quadratic equation parameterized by $\varphi$ as follows

$$at^2 + bt + c = 0$$
$$a = 1 \quad b = 2 s^T \xi(\varphi) \quad (56)$$
$$c = \xi^T(\varphi)\xi(\varphi) - r^2$$

i.e.

$$t^2 + 2 s^T \xi(\varphi) t + \xi^T(\varphi)\xi(\varphi) - r^2 \quad (57)$$
$$= 0$$

where

$$\xi(\varphi) = x_A - x_s(\varphi)$$
$$= \begin{bmatrix} x_A \\ y_A \\ z_C \end{bmatrix} - R \begin{bmatrix} cos\varphi \\ 0 \\ sin\varphi \end{bmatrix} \quad (58)$$

and

$$s = [s_x, s_y, 0]^T \quad (59)$$

If the Vieta's formula is used we get the following equivalent equations

$$t_1 + t_2 = -\frac{b}{a} = -2 s^T \xi(\varphi)$$
$$t_1 t_2 = \frac{c}{a} = \xi^T(\varphi)\xi(\varphi) - r^2$$

If a quadratic equation is considered as a quadratic function $F(t)$ of $t$, then the extreme value $t_{extrem}$ is given as

$$t_{extrem}(\varphi) = \frac{t_1 + t_2}{2} = -s^T \xi(\varphi) = \quad (60)$$

$$-[s_x, s_y, 0]^T \left( \begin{bmatrix} x_A \\ y_A \\ z_C \end{bmatrix} - R \begin{bmatrix} cos\varphi \\ 0 \\ sin\varphi \end{bmatrix} \right)$$
$$= s_x(R cos\varphi - x_A)$$

The point $x(t_{extrem}(\varphi))$ is inside of the envelope; see Fig.4, if and only if $F(t_{extrem}(\varphi)) < 0$. Substituting $t_{extrem}$ to the function $F(t)$

$$t^2 + 2s^T \xi(\varphi) t + \xi^T(\varphi) \xi(\varphi) - r^2 = 0 \quad (61)$$

we get

$$\begin{aligned}(-s^T \xi(\varphi))^2 \\ + (-s^T \xi(\varphi) . 2s^T \xi(\varphi)) \\ + \xi^T(\varphi) \xi(\varphi) - r^2 = 0\end{aligned} \quad (62)$$

i.e.

$$-s^T \xi(\varphi) + \xi^T(\varphi) \xi(\varphi) - r^2 = 0 \quad (63)$$

Substituting

$$\begin{aligned}s &= [s_x, s_y, 0]^T \\ \xi(\varphi) &= x_A - x_s(\varphi)\end{aligned} \quad (64)$$

This leads to:

$$\xi^T(\varphi) \xi(\varphi) - (s^T \xi(\varphi))^2 = r^2 \quad (65)$$

Therefore

$$\xi^T(\varphi)[I - s^T \otimes s]\xi(\varphi) = 0 \quad (66)$$

where $I$ is an identity matrix and $\otimes$ is a tensor product producing a matrix.

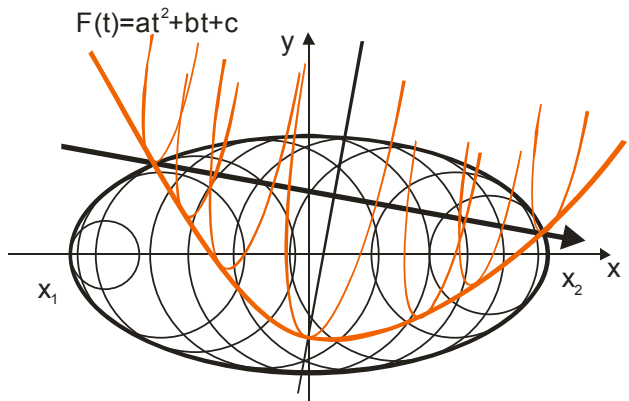

Figure 4: Rotating sphere plane intersection and envelope

As we recently set $a = 1$ in the quadratic equation, we can write

$$t_1(\varphi) t_2(\varphi) = c(\varphi) \quad (67)$$

where $t_1(\varphi)$ and $t_2(\varphi)$ are the line parameter values for line sphere intersection.

The second Vieta's [2] equation can be used to determine intervals for φ with one root only for iterative solvers.

In the classified case:
- ad a) we can use mirroring operations and solve the intersection in one quadrant only twice for non-mirrored and for mirrored cases as there might be two tuples of intersections,
- ad b) situation is complex as the envelope has an inflection point so there might be three intersections in one quadrant
- ad c) this case is similar to the previous but only two intersection points might occur

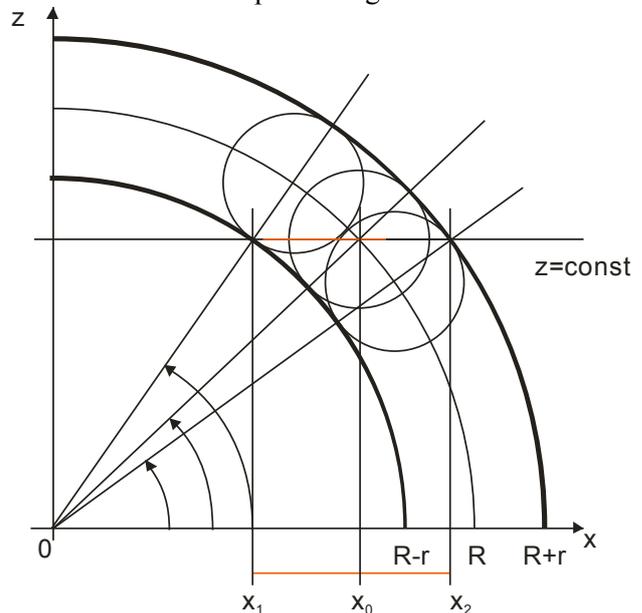

Figure 5: Rotating spheres

However the intersection computation is still too complex.

### 3.2 Line Rotation – Intersection in E³

Another alternative approach is based on a fixed sphere position on the $x$ axis and the given line rotates about $y$ axis generally in E³. This approach is actually "dual" in some sense to the previous one and leads to an envelope given as an intersection of a sphere and double cone.

There are two possible equivalent formulations: the center of a sphere is on the $x$ axis and the rotating line is in a general position in E³ or geometric transformation is made so that the rotating line rotates about $y$ axis and the vertex of a double cone is in the origin of the coordinate system; the center of a sphere is in the $x - y$ plane, i.e. was moved up.

A line in E³ is defined as
$$\begin{aligned}x(t) &= x_A + s\,t \\ s &= [s_x, s_y, 0]^T\end{aligned} \quad (68)$$

and a sphere on the $x$ axis is defined as
$$\begin{aligned}(x - x_s)^T (x - x_s) - r^2 &= 0 \\ x_s &= [R, 0, 0]^T\end{aligned} \quad (69)$$

As the line is rotated about y axis the rotation matrix $R$ is expressed as

$$R(\varphi) = \begin{bmatrix} \cos\varphi & 0 & -\sin\varphi \\ 0 & 1 & 0 \\ \sin\varphi & 0 & \cos\varphi \end{bmatrix} \quad (70)$$

Then the rotating line forming a double cone in $E^3$ can be expressed as

$$x(t,\varphi) = R(\varphi)(x_A + s\,t) \quad (71)$$
$$R(\varphi)x_A + R(\varphi)s\,t$$

Substituting we get

$$(R(\varphi)\,x_A + R(\varphi)\,s\,t - x_s)^T \\ (R(\varphi)\,x_A + R(\varphi)\,s\,t - x_s) - r^2 = 0 \quad (72)$$

or

$$(R(\varphi)\,s\,t + \xi)^T (R(\varphi)\,s\,t + \xi) - r^2 = 0 \quad (73)$$
$$\xi(\varphi) = R(\varphi)\,x_A - x_s$$

It means that a quadratic equation is obtained again, i.e.

$$s^T R^T(\varphi)\,R(\varphi)\,s t^2 \\ + 2s^T R^T(\varphi)\,\xi(\varphi) + \xi^T(\varphi)\xi(\varphi) \\ = 0 \quad (74)$$

As the matrix $R(\varphi)$ is orthonormal, i.e. $R^T(\varphi)\,R(\varphi) = I$ and directional vector can be normalized, i.e. $s^T s = 1$ then we get a significant simplification

$$t^2 + 2s^T R^T(\varphi)\xi(\varphi)t \\ + \xi^T(\varphi)\xi(\varphi) = 0 \quad (75)$$

Let us explore coefficients of this quadratic equation more in a detail.

$$s^T R^T(\varphi)\xi(\varphi) \\ = s^T R^T(\varphi)(R(\varphi)x_A - x_s) \\ = s^T R^T(\varphi)R(\varphi)x_A - s^T R^T(\varphi)x_s \quad (76)$$

As $R^T(\varphi)\,R(\varphi) = I$ we get

$$s^T R^T(\varphi)\xi(\varphi) = \\ s^T x_A - s^T R^T(\varphi)x_s \quad (77)$$

Using cross product symmetry we get

$$s^T R^T(\varphi)\xi(\varphi) = \\ s^T x_A - x_s^T R(\varphi)x_s \quad (78)$$

Now there is another simplification possible as $x_s = [R, 0, 0]^T$ and $s = [s_x, s_y, 0]^T$

$$s^T x_A - x_s^T R(\varphi)x_s \\ = [s_x, s_y, 0][x_A, y_A, z_A]^T \\ - [R, 0, 0] \begin{bmatrix} \cos\varphi & 0 & -\sin\varphi \\ 0 & 1 & 0 \\ \sin\varphi & 0 & \cos\varphi \end{bmatrix} \begin{bmatrix} R \\ 0 \\ 0 \end{bmatrix} \quad (79)$$
$$= s_x x_A + s_y y_A - R^2 \cos\varphi$$

Now the last term of the equation

$$\xi^T(\varphi)\xi(\varphi) = \\ [R(\varphi)x_A - x_s]^T[R(\varphi)x_A - x_s] \\ = x_A^T R^T(\varphi)R(\varphi)x_A \\ - 2x_A^T R^T(\varphi)x_s + x_s^T x_s \quad (80)$$

As $R^T(\varphi)\,R(\varphi) = I$

$$\xi^T(\varphi)\xi(\varphi) = \\ = x_A^T x_A - x_A^T R^T(\varphi)x_s + x_s^T x_s \\ = x_A^T x_A - x_s^T R(\varphi)x_A + x_s^T x_s \quad (81)$$

Using cross product symmetry we get

$$\xi^T(\varphi)\xi(\varphi) = \\ x_A^T x_A - 2x_s^T R(\varphi)x_A + x_s^T x_s \quad (82)$$

Again, there is another simplification possible as $x_s = [R, 0, 0]^T$ and $s = [s_x, s_y, 0]^T$

$$\xi^T(\varphi)\xi(\varphi) = \\ x_A^T x_A - 2x_s^T R(\varphi)x_A + x_s^T x_s \\ = x_A^T x_A \\ - 2[R, 0, 0]\begin{bmatrix} \cos\varphi & 0 & -\sin\varphi \\ 0 & 1 & 0 \\ \sin\varphi & 0 & \cos\varphi \end{bmatrix}\begin{bmatrix} x_A \\ y_A \\ z_A \end{bmatrix} \quad (83)$$
$$+ R^2$$
$$= x_A^T x_A - 2R\,(x_A\cos\varphi - z_A\sin\varphi) + R^2$$

Putting all together we get

$$t^2 + 2(s^T x_A - x_s^T R(\varphi)x_s)t \\ - 2x_s^T R(\varphi)x_A + x_A^T x_A + x_s^T x_s = 0 \quad (84)$$

i.e.

$$t^2 + (s_x x_A + s_y y_A - R^2\cos\varphi)t + x_A^T x_A \\ - 2R(x_A\cos\varphi \\ - z_A\sin\varphi) + R^2 = 0 \quad (85)$$

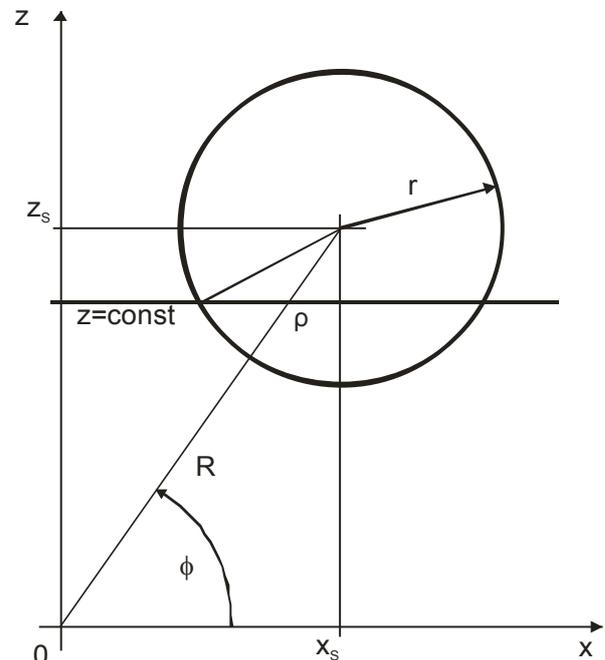

Figure 6: Intersection plane-rotating sphere

### 3.3 Intersection with a Plane - Solution in $E^2$

It this part we will concentrate on the case, when sphere rotates about $y$ axis and intersect a plane on which the given line lies and is parallel to the $x - z$ plane

As the given line lies in a plane parallel to the $x - z$ plane the rotating sphere intersect the plane, Fig.5, which results into circles in the $x - z$ plane, Fig.6.

$$(\pmb{x} - \pmb{x}_s)^T(\pmb{x} - \pmb{x}_s) - \varrho^2 = 0 \quad (86)$$

Let us consider the line formulation.

$$\begin{aligned}(\pmb{x} - \pmb{x}_s(\varphi))^T(\pmb{x} - \pmb{x}_s(\varphi)) - r^2 = 0 \\ \pmb{x}_s(\varphi) = R[cos\varphi, 0, sin\varphi]^T\end{aligned} \quad (87)$$

A sphere is rotating about $y$ axis is described by i.e.

$$\begin{aligned}[x - x_s(\varphi), y, z - z_s(\varphi)]^T \\ [x - x_s(\varphi), y, z - z_s(\varphi)] - r^2 = 0\end{aligned} \quad (88)$$

A plane on which the given line lies is defined as $z = z_c$. Then

$$\begin{aligned}(x - x_s)^2 + y^2 + (z - z_s)^2 - r^2 \\ = x^2 - 2xRcos(\varphi) + R^2cos^2(\varphi) \\ + y^2 \\ + z_c^2 - 2z_cRsin(\varphi) + R^2sin^2(\varphi) \\ - r^2 = 0\end{aligned} \quad (89)$$

As $cos^2(\varphi) + sin^2(\varphi) = 1$ we get

$$\begin{aligned}(x - x_s)^2 + y^2 + (z - z_s)^2 - r^2 \\ = x^2 + y^2 - 2R(xcos(\varphi) \\ + z_c sin(\varphi)) \\ + z_c^2 + R^2 - r^2 = 0\end{aligned} \quad (90)$$

As the given line is defined as

$$\begin{aligned}x(t) = x_A + s_x t \\ y(t) = y_A + s_y t\end{aligned} \quad (91)$$

we get

$$\begin{aligned}x_A^2 + 2x_A s_x t + s_x^2 t^2 + \\ y_A^2 + 2y_A s_y t + s_y^2 t^2 \\ -2R((x_A + s_x t)cos(\varphi) + z_c sin(\varphi)) \\ + z_c^2 + R^2 - r^2 = 0\end{aligned} \quad (92)$$

i.e. a quadratic equation has a form

$$\begin{aligned}(s_x^2 + s_y^2)t^2 \\ +2[(x_A - Rcos(\varphi))s_x + y_A s_y]t \\ -2R(x_A cos(\varphi) + z_c sin(\varphi)) \\ +x_A^2 + y_A^2 + z_c^2 + R^2 - r^2 = 0\end{aligned} \quad (93)$$

In the case of the normalized directional vector $\pmb{s}$, i.e. $\|\pmb{s}\| = 1$, resp. $\pmb{s}^T\pmb{s} = 1$, we get a quadratic equation parameterized by $\varphi$ as follows

$$\begin{aligned}t^2 + bt + c = 0 \\ b = 2[(x_A - Rcos(\varphi))s_x + y_A s_y] \\ c = -2R(x_A cos(\varphi) + z_c sin(\varphi)) \\ + x_A^2 + y_A^2 + z_c^2 + R^2 - r^2\end{aligned} \quad (94)$$

### 3.4 Hybrid method

Let torus is represented as an envelope of rotating spheres about $y$ axis again. Spheres intersect the plane $z = z_c$ on which the given line lies and form circles in the plane $z = z_c$, on the plane parallel to $z - y$ plane. Those circles on the plane are described by an equation

As all the circles are on the plane $z = z_c$ the equation can be simplified to

$$(x - x_s)^2 + y^2 - \varrho^2 = 0 \quad (95)$$

where

$$x_s = Rcos\varphi \quad y_s = 0 \quad z_s = Rsin\varphi \quad (96)$$

Note that $z_s$ represents rotation of the sphere about $y$ axis, resulting circle is on the $z = z_c$ plane. The $\varrho$ radius of a circle is given

$$\begin{aligned}\varrho^2 = r^2 - (z_s - z_c)^2 = \\ r^2 - (Rsin\varphi - z_c)^2\end{aligned} \quad (97)$$

The envelope of a plane-torus intersection is given as

$$\bigcup_{\varphi \in <\varphi_1, \varphi_2>} \{(x - Rcos\varphi)^2 + y^2 - r^2 \\ + (Rsin\varphi - z_c)^2 = 0\} \quad (98)$$

Let us consider the case, when $z_c < R - r$, Fig.7.

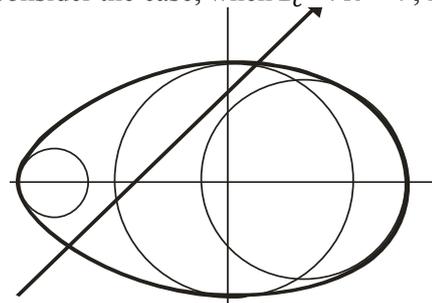

Figure 7: An envelope given as union of plane-rotating sphere intersections

Angles are determined as follows

$$\begin{aligned}cos(\varphi_1) = (R - r)^2 - z_c^2 \\ cos(\varphi_2) = (R + r)^2 - z_c^2 \\ cos(\varphi_0) = R^2 - z_c^2\end{aligned} \quad (99)$$

The angle $\varphi_1$ is an angle when the first circle that contributes to an envelope; the angle $\varphi_2$ is for the last circle that contributes to the envelope and the angle $\varphi_0$ is for the largest circle inside the envelope.

The given line lies in the $z = z_c$ plane and is defined as

$$\begin{aligned}x(t) = x_A + s_x t \\ y(t) = y_A + s_y t\end{aligned} \quad (100)$$

The line can be re-parameterized so that $y_A = 0$ then circles are defined as:

$$\begin{aligned}x(x_A + s_x t - Rcos\varphi)^2 + s_y^2 t^2 \\ -r^2 + (Rsin\varphi - z_c)^2 = 0\end{aligned} \quad (101)$$

Now the problem is effectively transferred to $E^2$.

### 3.5 New Bounding Volume

The "standard" bounding volume [1] is based on a sphere in $E^3$ and an intersection of two half spaces, Fig.2. As the line lies in the $x - y$ plane for $z = z_c$ we can distinguish following fundamental cases:
- ad a) we can use mirroring operations and solve the intersection in one quadrant only twice for non-mirrored and for mirrored cases as there might be two tuples of intersections,
- ad b) situation is complex as the envelope has an inflection point so there might be three intersections in one quadrant,
- ad c) this case is similar to the previous but only two intersection points might occur.

However if many lines-torus intersections computation are needed, like in the ray tracing rendering technique, the more precise bounding volume is needed to increase the efficiency of computation. The "standard" bounding volume works fine for the case ad b). On the other hand it can be seen that

- in the case ad a), i.e. when a line passes through the torus, i.e. through a "hole" and does not intersect the torus, detailed computation has to be made, that is computationally expensive.
- in the case ad c), i.e. when a line intersects the torus in its "outer part", i.e. $R \leq z_c < R + r$ the distance between two planes can be smaller than $2r$.

Let us explore the first case as it leads to higher efficiency.

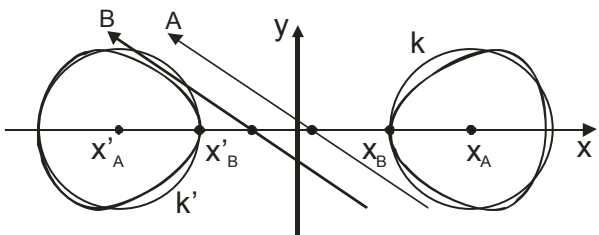

Figure 8: Torus-plane intersection and a ray

Fig.8 presents an intersection plane-torus for $0 < z_c < R - r$. It can be seen that a $k$ circle (as we are in $E^2$), with the center at $x_A$ with the radius $r$ forms bounding surfaces together with the mirrored $k'$ circle by $y$ axis. The $x_A$ center of the circle is defined as follows:

$$x_B = (R - r)\cos\varphi$$
$$x_A = x_B + r = (R - r)\cos\varphi + r \quad (102)$$

where

$$\sin\varphi = \frac{z_c}{R} \quad (103)$$

or

$$x_B = \sqrt{(R - r)^2 - z_c^2}$$
$$x_A = x_B + r = r + \sqrt{(R - r)^2 - z_c^2} \quad (104)$$

It can be seen that in the case of $0 < z_c = R - r$ a special case is obtained as there is no "hole" at all, Fig.9

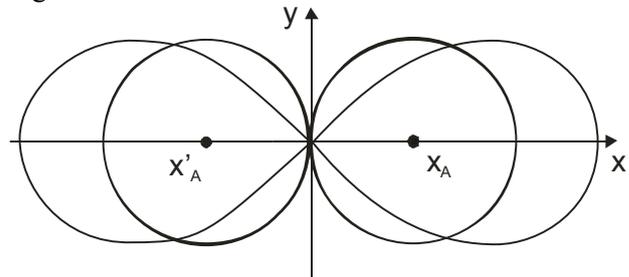

Figure 9: A boundary situation

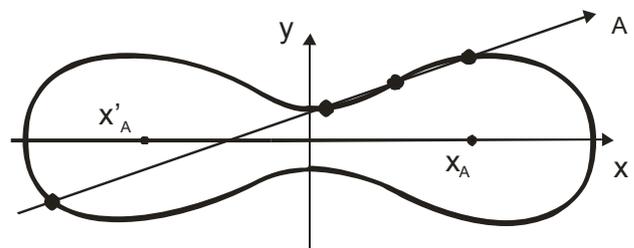

Figure 10: Line-torus intersection for $R - r \leq z_c < R$, i.e. the case ad b)

The test for the ad a) case can be formulated as: if the line intersects the $x$ axis in the interval $(x'_B, x_B)$ and does not intersect the circle $k$ nor the circle $k'$, then the line does not intersect the given torus. Fig.6 presents two lines, in the case A, the line is not considered for intersection computation with torus, while in the cases B, the detailed intersection test/computation has to be made.

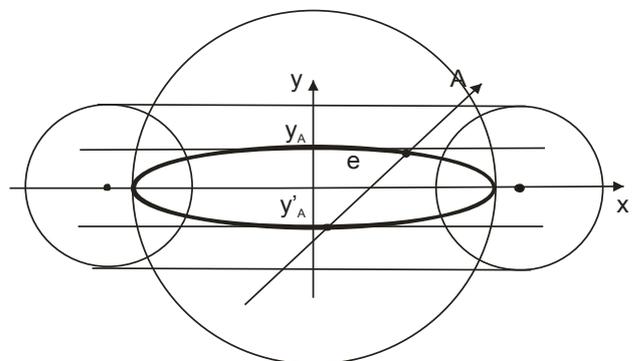

Figure 11: Line-torus intersection and bounding for $R \leq z_c < R + r$, i.e. the case ad c)

The test for the ad b) test remains as the original, Fig.10, as up to 3 intersections can occur in one quadrant as there is a point of inflexion.

In the case ad c), i.e. $R \leq z_c < R + r$, there are only 2 intersection points possible, Fig.11. It can be seen that the distance between two planes, given by $y_A$ and $y'_A$ values is now smaller than the original distance $2r$. It can be seen that the new distance is given as

$$d^2 = r^2 - (z_c - r)^2 \qquad (105)$$

## 4 Conclusion

New alternative formulations for line-torus intersection problem have been presented. Unfortunately all the presented alternative formulations do not lead to simpler computational formulas. It seems to that an implicit form for the line-torus intersection is the most efficient one. There is still one possibility to use toroidal coordinate system; however the computational expense is too high.

As a result of new geometrically equivalent formulations a new bounding object, actually circles in $E^2$, for the line-torus intersection has been developed and described.

The new bounding object increases line-torus intersection computation efficiency significantly as it also detects the cases when a line or ray is passing a "hole" of the torus. The efficiency of the new torus bounding test grows with the ratio $v = R/r$.

## Acknowledgment

The author would like to thank to colleagues and students at the VSB-Technical University of Ostrava and University of West Bohemia for their critical comments, suggestions and hints. Thanks belong also to anonymous reviewers for corrections and comments.

This research was supported by the Ministry of Education of the Czech Republic, projects No.LH12181, LG13047.